\documentclass[14pt,a4paper]{extarticle}

\usepackage[utf8]{inputenc}
\usepackage[english]{babel}
\usepackage[T2A]{fontenc}

\usepackage{amsmath}
\usepackage{amsfonts}
\usepackage{amssymb}
\usepackage{graphicx}
\usepackage{hyperref}
\usepackage{color}
\usepackage{fullpage}
\usepackage{indentfirst}
\usepackage{tabularx}
\usepackage{cite}

\definecolor{darkblue}{RGB}{23,26,73}
\definecolor{darkgreen}{RGB}{27,37,14}
\definecolor{darkred}{RGB}{37,14,14}

\newcommand{\rr}{{\vec{r}}}

\hypersetup{%
  colorlinks=true,%
  linkcolor=darkred,
  citecolor=darkgreen,
  urlcolor=darkblue,
  pdfauthor={A. A. Lagutin, and N. V. Volkov},%
  pdftitle={Nonmonotonic change with energy of the mean logarithmic mass of cosmic rays in the knee region:
the mechanism of formation of this feature and sources of particles},%
  pdfsubject={Manuscript of an article for the journal PHAN},%
  pdfkeywords={Cosmic rays, energy spectrum, mean logarithmic mass, knee region, PeVatrons, non-classical diffusion model, astrophysical interpretation},%
  breaklinks=true,%
  plainpages=false%
}

\begin{document}

\begin{center}
{\large\bf Nonmonotonic change with energy of the mean logarithmic mass of cosmic rays in the knee region:
the mechanism of formation of this feature and sources of particles}

\vspace*{1cm}

A. A. Lagutin\footnote{\url{https://orcid.org/0000-0002-1814-8041}} and N. V. Volkov\footnote{\url{https://orcid.org/0000-0002-3172-0655}}

\vspace*{5mm}

Altai State University, Barnaul, 656049 Russia

\vspace*{5mm}

e-mail: \verb|lagutin@theory.asu.ru|

\end{center}

\begin{abstract}

Recently, the Large High Altitude Air Shower Observatory (LHAASO) published measurements of the all-particle CR energy spectrum and the mean logarithmic mass of CRs with unprecedented accuracy in 0.3$-$30~PeV. The mean logarithmic mass shows a nonmonotonic change with energy, a feature observed for the first time. In this work, we present a new approach to describe the mechanisms of formation of this feature. The key elements of this approach are the non-classical diffusion model of cosmic rays developed by the authors in which the knee in the observed spectrum occurs naturally without the use of additional assumptions, as well as power-law asymptotics before and after the knee, and a soft spectrum of particle generation in cosmic ray source. To obtain a more complete picture of the spectrum formation in the region of the knee and the sources that form it, we carried out calculations of the spectra of the main groups of nuclei in the energy range of 1~TeV$-$100~PeV. It is shown that the behavior of the all-particle spectrum and mass composition in the knee region is determined by local pevatrons located at a distance of 750$-$900~pc from the Earth. The energy of the knee practically coincides with the knees in the spectra of protons and helium nuclei. The contribution of the light components $p+$He is about 70\%, the CNO group provides $\sim$13\%. The energy spectrum index of the light components is $-2.61$ before the knee. The nonmonotonic change in the mean logarithmic mass is due mainly to a decrease in the contribution of the CNO group in the energy range of 0.3$-$3~PeV.

\vspace*{5mm}

\textbf{Keywords:} cosmic rays, energy spectrum, mean logarithmic mass, knee region, PeVatrons, non-classical diffusion model, astrophysical interpretation.

\end{abstract}

\section*{Introduction}

The break in the spectrum of cosmic rays (CR) at the energy of $\sim 3$~PeV (the so-called <<knee>>), first observed in its electromagnetic component~\cite{Kulikov:1958}, is the most expressive feature around which numerous studies aimed at solving the problem of the CRs origin are built. In the standard paradigm, the knee is explained as a superposition of cutoffs in the spectra of protons and other groups of nuclei~\cite{Lagage:1983}. The mechanism of acceleration CRs by diffusive shock waves and diffusive propagation in the Galaxy are the fundamental statements of this paradigm~\cite{Blasi:2013}. At the same time, the experimental results of recent years decisively indicate the need to revise some of its provisions.

Latest key results in the energy region around the knee were obtained by the LHAASO observatory. The paper~\cite{lhaaso:2025all} publishes measurements of the all-particle CR energy spectrum and the mean logarithm of the mass number, obtained with unprecedented accuracy in the range of 0.3$-$30~PeV. One of the main results~\cite{lhaaso:2025all} is the nonmonotonic change in the mean logarithmic mass with energy, a feature observed for the first time. As a result of the analysis of the latest LHAASO data on high-precision measurements of the proton spectrum in the energy range of 0.15$-$12~PeV~\cite{lhaaso:2025p}, it was established that the approximation of experimental data by a combination of power-law functions is more preferable than the model with an exponential cutoff, which is often used in the interpretation of energy spectra taking into account the maximum energy threshold for particle acceleration in sources. Thus, this clearly indicates that CR sources effectively accelerate protons and, apparently, other nuclei to energies greater than those accepted as the limit of the CRs spectrum of galactic origin. In the paper~\cite{lhaaso:2025p} these sources are called <<PeVatrons>> or even <<super-PeVatrons>>.

Another confirmation of effective acceleration in galactic sources is the recent results of the H.E.S.S observatory~\cite{hess:2024} on the spectrum of the leptonic component in the region of 0.3$-$40~TeV, which show the presence of a break in the total spectrum of electrons and positrons at the energy of $\sim 1$~TeV, as well as power-law asymptotics of the spectrum before and after the break. At $E>1$~TeV the spectrum extends far beyond the limits of direct measurements by DAMPE~\cite{dampe:2017}, CALET~\cite{calet:2023} and other observatories, which also indicates effective acceleration of electrons and positrons in galactic sources to energies greater than those previously considered threshold.

Finally, the registration of gamma quanta with PeV energies~\cite{asgamma:2021,lhaaso:2021gamma} indicates that, regardless of the channel of their origin (lepton or pion), the particles that emitted them have energies greater than the conventional threshold values. Today LHAASO has discovered 43 ultra-high-energy gamma-ray sources in the
Galaxy with the maximal energy of photons reaching up to 2.5 PeV~\cite{lhaaso:2024gamma}.

Phenomenological studies of the cosmic ray flux and its mass composition in the knee region have been conducted in many papers utilizing the most up-to-date data available at different times~\cite{horandel:2003,zatsepin:2006,gaisser:2013,thoudam:2016,dembinski:2018}. However, the technology underlying these studies is not a model in the usual sense, since it does not attempt to explain the data and, as a result, does not allow the parameters of the sources to be reconstructed.

The main goal of this paper is to formulate a new approach to description of the mechanisms of the features formation of the CRs spectrum and mass composition in the knee region. The key element of this approach is the model of non-classical CR diffusion developed by the authors, in which the break in the observed spectrum occurs naturally without using additional assumptions. Power-law asymptotics before and after the break, as well as the soft spectrum of particle generation in CR sources are the other key properties of the model. These properties of the model are the basis of new tools for reconstructing the parameters of sources that form the knees of the spectra.

\section{Non-classical diffusion model}

The features of the energy spectra of nuclei, electrons and positrons established in the experiments ATIC-2~\cite{atic-2:2009}, PAMELA~\cite{pamela:2013}, Fermi-LAT~\cite{fermi:2017}, H.E.S.S~\cite{hess:2024}, AMS-02~\cite{ams-p:2015,ams-he:2015}, NUCLEON~\cite{NUCLEON:2019}, DAMPE~\cite{dampe:2017,dampe-p:2019,dampe-he:2021}, CALET~\cite{calet:2023,calet-p:2022,calet-he:2023}, LHAASO~\cite{lhaaso:2025all,lhaaso:2025p} and others, stimulated numerous studies aimed at interpreting the obtained data. For example, in the papers~\cite{Yue:2020,Malkov:2021,Li:2024,Panov:2024,Bhadra:2025} (see also references in these papers) the identified spectral features were discussed within the framework of the standard scenario of cosmic rays propagation in the interstellar medium. In our papers~\cite{lagutin-bras:2021,lagutin-phan:2021,lagutin-bras:2023,lagutin-phan:2023,lagutin-bras:2025} an original scenario was formulated that allows for a clear astrophysical interpretation of the revealed complex structure of the energy spectra of particles. The key element of this scenario is the non-classical diffusion model of cosmic rays developed by the authors.

The non-classical diffusion model, first proposed in the works~\cite{Lagutin:2001CRC,Lagutin:2001NP,Lagutin:2003}, is a generalization of the Ginzburg-Syrovatskii normal diffusion model~\cite{Ginzburg:1964} for the case of particle propagation in a highly inhomogeneous galactic medium of a fractal type. The need for such a generalization is due to the presence in the interstellar medium of structures with complex morphology (filaments, ribbons, clouds and voids)~\cite{Elmegreen:1996,Heyer:1998,Elmegreen:2004,Bergin:2007,Sanchez:2008,Efremov:2003,Fuente:2009,Sanchez:2010} and an intermittent magnetic field~\cite{Zeldovich:1990}. A consequence of the irregular distribution of matter and the magnetic field associated with it is the power-law distribution of the free paths of particles $r$ between inhomogeneities $p(r) \propto A(E,\alpha) r^{-\alpha - 1}, r \rightarrow \infty, 0 < \alpha < 2$ (L\'{e}vy flights), as well as the power-law distribution of the time of residence of particles $t$ in inhomogeneities (L\'{e}vy traps) $q(t) \propto B(E,\beta)t^{-\beta - 1}$, $t \rightarrow \infty$, \mbox{$\beta < 1$}.

The generalized non-classical diffusion equation for the particles density with energy $E$, injected at point $\rr$ at time $t$ by the distribution of sources density $S(\rr,t,E)$ without taking into account energy losses and nuclear interactions of cosmic rays, was first obtained in the works~\cite{Lagutin:2001CRC,Lagutin:2003} and has the form
\begin{equation}\label{eq:nonclass-diff-eq}
\frac{\partial N(\rr, t, E)}{\partial t}= -D(E, \alpha, \beta)\mathrm{D}_{0+}^{1-\beta}(-\Delta)^{\alpha/2} N(\rr, t, E) + S(\rr, t, E).
\end{equation}
Here $D(E, \alpha, \beta) = D_0(\alpha, \beta) \left(E/1~\text{GeV}\right)^{\delta}$ is the anomalous diffusivity. From the physical point of view, the fractional Laplacian (<<Riesz operator>>)\linebreak $(-\Delta)^{\alpha/2}$~\cite{Samko:1993} and the fractional Riemann-Liouville derivative $\mathrm{D}_{0+}^{\beta}$~\cite{Samko:1993} in Eq.~\eqref{eq:nonclass-diff-eq} reflect the nonlocality and non-Markovian nature of the diffusion process, respectively. Note that for $\alpha=2$, $\beta=1$ from~\eqref{eq:nonclass-diff-eq} we obtain the Ginzburg-Syrovatskii normal diffusion equation~\cite{Ginzburg:1964}.

The equation for the Green's function $G(\rr, t, E; \rr_0, t_0, E_0)$, which should be interpreted as the probability of detecting a particle at the point $(\rr, t, E)$ of the phase space if at the initial moment its characteristics were $(\rr_0 , t_0, E_0)$, we write in the form~\cite{Lagutin:2001CRC,Lagutin:2003}
\begin{multline}\label{eq:green-func}
\frac{\partial G(\rr, t, E; \rr_0, t_0, E_0)}{\partial t}=-D(E, \alpha, \beta)\mathrm{D}_{0+}^{1-\beta}(-\Delta)^{\alpha/2} G(\rr, t, E; \rr_0, t_0, E_0) +\\ 
+\delta(\rr-\rr_0)\delta(t-t_0)\delta(E-E_0).
\end{multline}
For $\rr_0 = 0$ and $t_0=0$, the solution to Eq.~\eqref{eq:green-func}, which we find using the Fourier-Laplace transforms of spatial and temporal coordinates, is the Green's function
\begin{equation}\label{eq:green-function}
G(\rr, t, E; E_0) = \delta(E-E_0)\left(D t^\beta\right)^{-3/\alpha}\Psi_3^{(\alpha, \beta)}\left(|\rr|\left(D
t^\beta\right)^{-1/\alpha}\right).
\end{equation}
Here
\begin{equation*}\label{eq:psi-stable-distribution}
\Psi_3^{(\alpha, \beta)}(r)=\int\limits_0^\infty g_3^{(\alpha)}\left(r\tau^{\beta/\alpha}\right) q_1^{(\beta)}(\tau) \tau^{3\beta/\alpha} d\tau
\end{equation*}
is the density of the fractional-stable distribution~\cite{Uchaikin:1999a} defined by the three-dimensional spherically symmetrical stable distribution $g_3^{(\alpha)}(r)$ (\mbox{$0<\alpha\leqslant 2$}) and the one-dimensional one-sided stable distribution $q_1^{(\beta)}(t)$ with the characteristic parameter $0<\beta\leqslant 1$. It should be noted that for $\alpha=2$ the density $g_3^{(\alpha)}(r)$ is the Gaussian distribution, and for $\beta=1$ the density $q_1^{(\beta)}(t)$ is the Dirac delta function. In this case, from~\eqref{eq:green-function} we obtain the well-known result of the normal diffusion model~\cite{Ginzburg:1964}.

The Green's function~\eqref{eq:green-function} allows us to find a solution of the non-classical diffusion equation~\eqref{eq:nonclass-diff-eq} for the cosmic ray density $N(\rr, t, E)$ in the case of sources described by the distribution $S(\rr, t, E)$. By definition, we have
\begin{equation*}\label{eq:NGS}
N(\rr, t, E) = \int\limits_{\mathrm{R}^3} d \rr_0 \int\limits_{-\infty}^t d t_0 \int\limits_E^{\infty} d E_0
G(\rr, t, E; \rr_0, t_0, E_0) S(\rr_0, t_0, E_0).
\end{equation*}
In the case of a point pulse source with a power-law energy spectrum of particle injection
\begin{gather*}\label{eq:point-impulse-source}
S(\rr, t, E) = S_{\text{0}} E^{-\gamma} \delta(\rr) H(T-t)H(t),\\
H(\tau) =
\begin{cases}
1,&\tau > 0,\\
0,&\tau < 0,\\
\end{cases}\notag
\end{gather*}
modeling the spectrum of particles accelerated in the source during a time interval $T$, the solution of the non-classical diffusion equation~\eqref{eq:nonclass-diff-eq} has the form~\cite{Lagutin:2001CRC,Lagutin:2003}
\begin{equation}\label{eq:point-impulse-concentration}
N(\rr, t, E)= \dfrac{S_{\text{0}} E^{-\gamma}}{D(E, \alpha, \beta)^{3/\alpha}}\int\limits_{\max[0,t-T]}^t d\tau \tau^{-3\beta/\alpha} \Psi_3^{(\alpha,\beta)}\left(|\rr|(D(E, \alpha, \beta)\tau^\beta)^{-1/\alpha}\right).
\end{equation}

The analysis of the behavior of $N(\rr, t, E)$ in the case of a point pulse source from energy $E$ was first performed in~\cite{Lagutin:2001CRC,Lagutin:2001NP,Lagutin:2003}. The authors showed the presence of a break in the spectrum, which is the most expressive observed feature of the CR spectrum. An important result of the non-classical diffusion model is the conclusion that at the break point at $E=E_k$ the exponent of the observed spectrum is equal to the particle generation spectrum exponent in the source $\gamma$, and the spectrum itself has power-law asymptotics before ($E\ll E_k$) and after ($E\gg E_k$) the break point of the following form
\begin{equation}\label{eq:assympt}
\begin{split}
N\sim E^{-\gamma + \delta}, E\ll E_k \\
N\sim E^{-\gamma - \delta/\beta}, E\gg E_k.
\end{split}
\end{equation}
The results of calculations of the observed spectrum exponent in the case of a point pulse source are shown in~\cite{lagutin-phan:2021}. In particular, in paper~\cite{lagutin-phan:2021} shows that the break in the spectrum occurs only in the non-classical diffusion regime with parameters $1<\alpha<2$, $\beta < 1$ and in the superdiffusion regime ($1<\alpha<2, \beta=1$). In the subdiffusion regime ($\alpha=2, \beta<1$) and the normal diffusion one ($\alpha=2, \beta=1$), the spectrum has no break.

Another important result was obtained within the framework of the stationary model of non-classical diffusion~\cite{Lagutin:2001s,Lagutin:2025asu}. The authors~\cite{Lagutin:2001s,Lagutin:2025asu} showed that in this case the spectrum also has no break and the observed spectrum is softer than the spectrum in the source by $\delta/\beta$, i.e.
\begin{equation}\label{eq:stac-superdiff-concentration}
  N(\rr, E)\sim E ^{-\gamma-\delta/\beta}.
\end{equation}

In conclusion of this section, we once again note the most important results of the non-classical CR diffusion model.
\begin{enumerate}
\item The presence of a break in the non-classical diffusion model for regime with $1<\alpha<2$ and $\beta \leq 1$ is an internal property of the model that does not require additional assumptions.
\item The power-law asymptotics of the spectra of individual groups of nuclei, as well as the spectrum of all particles, established in DAMPE~\cite{dampe-p:2019,dampe-he:2021}, CALET~\cite{calet-p:2022,calet-he:2023}, LHAASO~\cite{lhaaso:2025all,lhaaso:2025p} and other experiments before and after the breaks are convincing evidence in favor of the non-classical nature of CR diffusion in the Galaxy.
\item The soft spectrum of particle generation in sources $\gamma > 2.5$, established in the non-classical diffusion model, is also confirmed in the experiments.
\end{enumerate}

\section{Energy spectrum}

Following another provision of the scenario proposed in~\cite{lagutin-bras:2021,lagutin-phan:2021,lagutin-bras:2023,lagutin-phan:2023,lagutin-bras:2025} for the interpretation of the observed CR spectra, in this paper, when calculating the energy spectra of the nuclear component, galactic CR sources are divided into two groups. The first group includes numerous old ($t\geq 10^6$~years) distant ($r\geq 1$~kpc) sources. The second group is nearest ($r< 1$~kpc), young ($t< 10^6$~years) sources. The spatial separation of the sources leads to the separation of the observed cosmic ray fluxes from them:
$$J = J_{\text{G}}(\rr, E) + J_{\text{L}}(\rr, t, E).$$

Here $J_{\text{G}}$~ is the global component of the spectrum, taking into account the contribution from the system of old distant stationary sources. This spectrum component is calculated using the results of the stationary non-classical diffusion model~\eqref{eq:stac-superdiff-concentration}. The local component of the spectrum $J_{\text{L}}(\rr, t, E)$ includes contributions from two groups of nearby young sources, which determine the hardening and the break in the spectrum of the light nuclear component in the TeV energy region (the so-called <<tevatrons>>) and the second hardening and knee in the region $E>10^5$~GeV (the so-called <<pevatrons>>). The contribution from these groups of sources is calculated using the expression~\eqref{eq:point-impulse-concentration} for the particle density from a point pulse source.

Thus, the observed CR flux in the energy range considered in this paper is equal to
$$J = J_{\text{G}}(\rr, E) + J_{\text{TeV}}(\rr,t,E) + J_{\text{PeV}}(\rr,t,E).$$

\begin{figure}[ht!]
\includegraphics[width=1.1\textwidth]{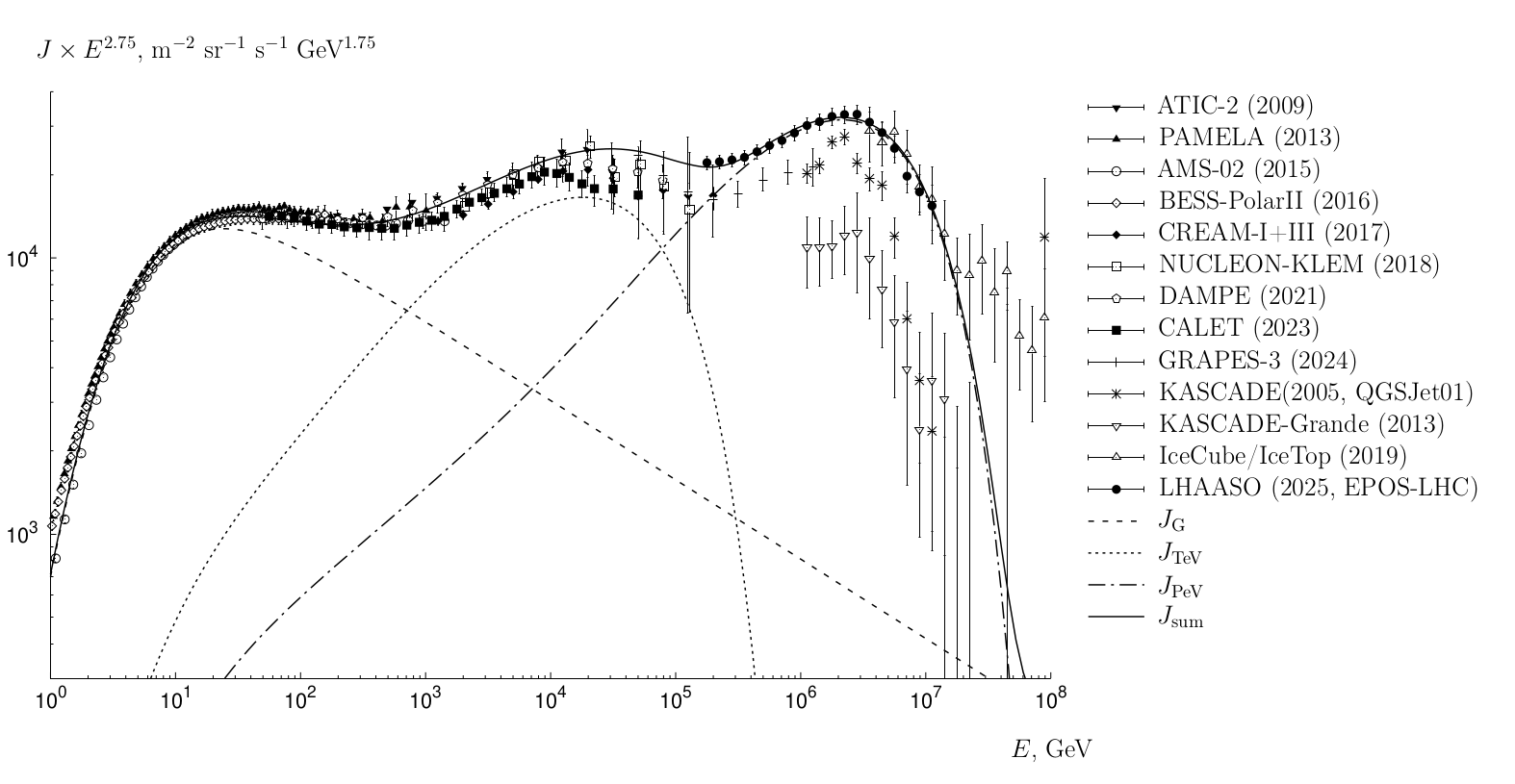}
\caption{Comparison of the proton spectrum obtained in the non-classical diffusion approach with the experimental data \mbox{ATIC-2} (2009)~\cite{atic-2:2009}, PAMELA (2013)~\cite{pamela:2013}, AMS-02 (2015)~\cite{ams-p:2015}, BESS-PolarII (2016)~\cite{BESS-Polar:2016}, CREAM-III (2017)~\cite{CREAM-III:2017}, NUCLEON (2019)~\cite{NUCLEON:2019}, DAMPE (2019)~\cite{dampe-p:2019}, CALET (2019)~\cite{calet-p:2022}, KASCADE (2005)~\cite{KASKADE:2005}, KASCADE-Grande (2013)~\cite{KASKADE-Grande:2013ap}, IceCube/IceTop (2019)~\cite{IceCube:2019}, LHAASO (2025)~\cite{lhaaso:2025p}}\label{fig:p-spectrum}
\end{figure}

Figure~\ref{fig:p-spectrum} shows the results of calculations of the proton spectrum within the framework of the non-classical diffusion model. The model parameters used in the calculations are given in Table~\ref{tab:adparams}. The technique of self-consistent parameter recovery is discussed in the papers~\cite{lagutin-bras:2021,lagutin-phan:2021,lagutin-bras:2023,lagutin-phan:2023,lagutin-bras:2025,Lagutin:2019}. The energy spectra of He, C, O are shown in~\cite{lagutin-phan:2021,lagutin-bras:2025}. For the other groups of nuclei, the calculations were carried out similarly.

\begin{center}
\begin{table}[hp]
\centering
\caption{The non-classical diffusion model parameters}\label{tab:adparams}
\begin{tabularx}{\textwidth}{|c|c|X|}
\hline
\textbf{Parameter} & \textbf{Value} & \textbf{Description} \\
\hline
$\gamma$ & 2.7 for $p$ and other nuclei, 2.6 for He & Energy spectral index of particles accelerated at source \\
\hline
$\delta$ & $0.27$ & Exponent of energy dependency of diffusion coefficient \\
\hline
$D_0(\alpha,\beta)$ & $ 1.5\times 10^{-3}$~pc$^{1.7}$yr$^{-0.8}$ & Anomalous diffusivity \\
\hline
$\alpha$ & $1.7$ & Exponent of the power-law distribution of the free paths of particles between inhomogeneities\\
\hline
$\beta$ & $0.8$ & Exponent of the power-law distribution of the time of particles residence in inhomogeneities\\
\hline
$T$ & $10^4$ yr & Time of injections of particles in sources \\
\hline
\end{tabularx}
\end{table}
\end{center}

\newpage

\section{Tools for restoring source parameters}

The unique properties of the non-classical diffusion model (the presence of a break, power-law asymptotics of the spectrum before and after the break, and the soft spectrum of particle generation in sources) make it possible to propose a tools for reconstructing source parameters (distance and age) from available experimental data. The calculation algorithm includes the following steps:
\begin{enumerate}
\item Assuming that the features of the spectrum in the energy range under consideration are mainly due to the contribution from one group of sources, we subtract from the experimental data the contributions from the remaining groups of sources and background, calculated within the framework of the non-classical diffusion model.

\item Next, we fit the transformed experimental data with power functions before and after the break. As a result, we obtain the spectrum indices that would be observed under the condition that the spectrum in the break energy region is formed by only one group of sources.

\item Using the obtained estimates and the theoretically established asymptotics of the spectrum~\eqref{eq:assympt}, we reconstruct the index of the particle generation spectrum $\gamma$ in the source and the exponent of the power-law distribution of the time of particles residence in inhomogeneities $\beta$.
\end{enumerate}

The proposed tools was used in our papers~\cite{lagutin-phan:2023,lagutin-bras:2025} to determine the parameters of tevatrons that leads to hardening and the break in the spectra of protons and helium nuclei in the TeV energy region. As a result of calculations, it was established that the breaks in the proton and helium spectra revealed in the DAMPE~\cite{dampe-p:2019,dampe-he:2021} and CALET~\cite{calet-p:2022,calet-he:2023} experiments can be explained by the contribution of a local group of sources located at a distance of $\sim$150$-$200~pc from the Earth and having an age of $\sim (4-6)\times 10^5$~years. These sources inject particles into the interstellar medium with indices of $-2.7$ for protons and $-2.6$ for helium nuclei.

\begin{figure}[ht!]
\begin{center}
\includegraphics[width=.75\textwidth]{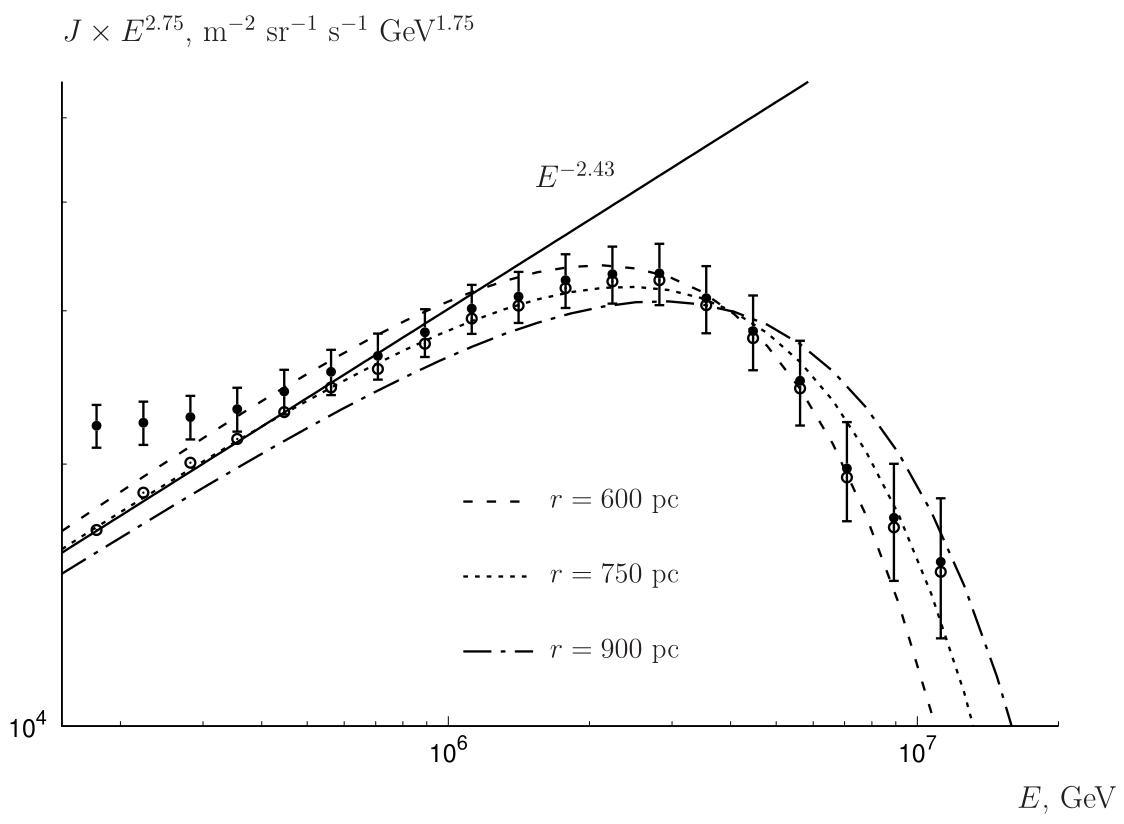}
\end{center}
\caption{Spectra of protons observed by LHAASO~\cite{lhaaso:2025p} (filled points). Open points are the local pevatron spectrum $J_{\text{PeV}}$, obtained within the proposed tools for reconstrution of parameters of sources after subtracting from LHAASO spectrum of the contributions of the both global-scale steady state component $J_{\text{G}}$ and tevatrons one $J_{\text{TeV}}$. Solid line on this figure is the power-law approximations of transformed LHAASO spectrum of pevatrons before the break. Other lines on this figure are shown the dependence of the proton spectrum from the distance to pevatrons $r$ having age equal to $t=0.5\times 10^5$~yr}\label{fig:lhaaso-spectrum}
\end{figure}

In this paper the proposed tools was used to analyze the LHAASO~\cite{lhaaso:2025p} data on the proton spectrum in the energy range of $10^5-10^7$~GeV. Figure~\ref{fig:lhaaso-spectrum} shows the original LHAASO data, as well as the data after subtracting the contribution of the global component $J_{\text{G}}$ and the contribution of the local group of tevatrons $J_{\text{TeV}}$. It was found that the knee in the proton spectrum at the energy of $E\sim3$~PeV can be explained by the contribution of another local group of sources (pevatrons) with a generation spectrum index of $-2.7$, located at a distance of $\sim$750$-$900~pc from the Earth. The age of pevatrons is $t<10^5$~years. Note that the parameters of the pevatrons obtained using the proposed tools are consistent with the results of our paper~\cite{lagutin-phan:2023pev}, in which the distance to pevatrons were obtained as a result of a joint analysis of the proton and electron spectra at the knees points.

\begin{center}
\begin{table}[h]
\centering
\caption{Space-time parameters of the most likely candidates (supernova remnants) for tevatrons and pevatrons according to~\cite{Kobayashi:2004,Lozinskaya:1992}. Parameters of the nearby star cluster RCW 36 ($r=900$~pc, $t=1.1$~Myr)~\cite{Peron:2024} also allow it to be classified as a pevatron candidate.}\label{tab:sources}
\begin{tabular}{|l|l|l||l|l|l|}
\hline
\multicolumn{3}{|c||}{Tevatrons} & \multicolumn{3}{c|}{Pevatrons} \\
\hline
Source & $r$, pc & $t, 10^5$~yr & Source & $r$, pc & $t, 10^5$~yr \\
\hline
Loop I & 120 & 2.0 & Monoceros & 600 & 0.46 \\
Loop II & 175 & 4.0 & Cyg. Loop & 770 & 0.20 \\
Loop III & 200 & 4.0 & CTB 13 & 600 & 0.32 \\
Loop IV & 210 & 4.0 & S 149 & 700 & 0.43 \\
Vela & 250 & 0.46 & STB 72 & 700 & 0.32 \\
Geminga & 400 & 3.4 & CTB 1 & 900 & 0.47 \\
Lopus Loop & 400 & 0.36 & HB 21 & 800 & 0.23 \\
\hline
\end{tabular}
\end{table}
\end{center}

Table~\ref{tab:sources} shows the space-time coordinates of the most likely candidates for tevatrons and pevatrons.

\section{Results of calculations}

\begin{figure}[ht!]
\begin{center}
\includegraphics[width=\textwidth]{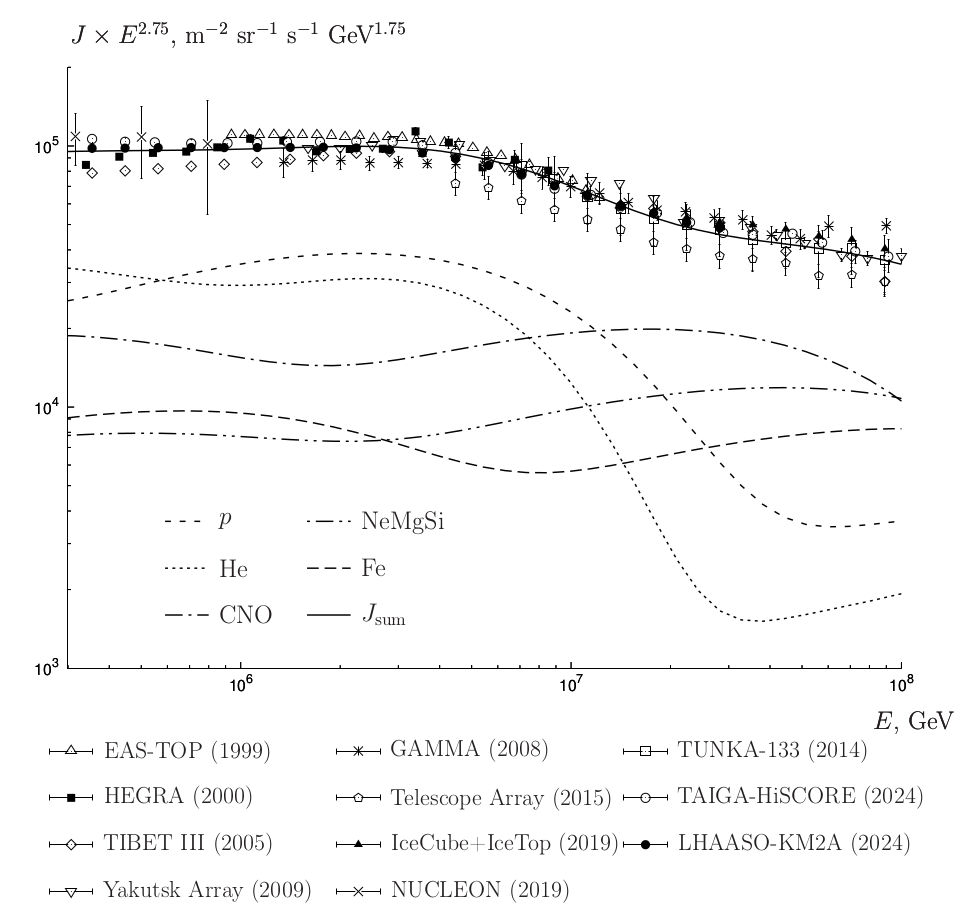}
\end{center}
\caption{Comparison of the all particles spectrum obtained in the non-classical diffusion approach with the experimental data EAS-TOP (1999)~\cite{eastop:1999}, HEGRA (2000)~\cite{HEGRA:2000}, TIBET III (2005)~\cite{TIBETIII:2008}, Yakutsk Array (2009)~\cite{Yakutsk:2009}, GAMMA (2008)~\cite{GAMMA:2008}, Telescope Array (2015)~\cite{TelescopeArray:2015}, IceCube+IceTop (2019)~\cite{IceCube:2019}, NUCLEON (2019)~\cite{NUCLEON:2019}, TUNKA-133 (2014)~\cite{TUNKA133:2014}, LHAASO-KM2A (2024)~\cite{lhaaso:2025all}, TAIGA-HiSCORE (2024)~\cite{taiga:2025}}\label{fig:all-spectrum}
\end{figure}

To obtain a complete picture of the CR spectrum formation in the region of the knee, the spectra of the main groups of nuclei were calculated in the energy range of 1~TeV$-$100~PeV. The figure~\ref{fig:all-spectrum} shows the results of comparing the all particles spectrum calculations, obtained within the framework of the non-classical diffusion model, with the experimental data. It can be seen that the knee position practically coincides with the knees in the spectra of protons and helium nuclei. Additional calculations show that the exponent of the energy spectrum of light component is equal to $-$2.61 before the knee. This estimate is in good agreement with the LHAASO results~\cite{lhaaso:2025all}. For the CNO, NeMgSi and Fe groups of nuclei, the energy range 1$-$10~PeV is a transition region between the contributions from tevatrons and pevatrons. A decrease in the contributions from these groups of nuclei was established in this region.

\begin{figure}[ht!]
\begin{center}
\includegraphics[width=\textwidth]{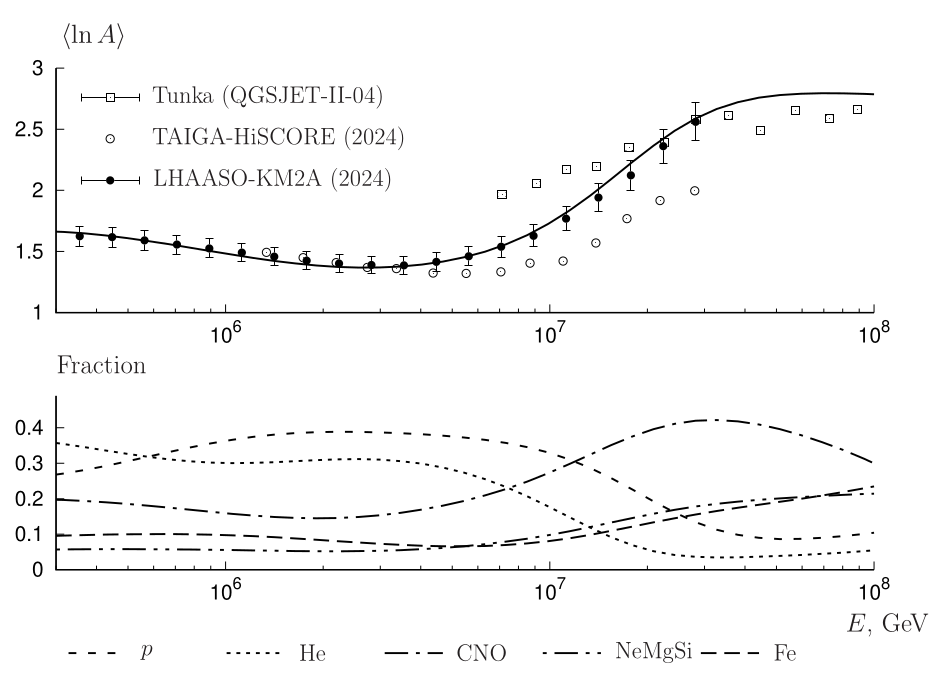}
\end{center}
\caption{The mean logarithm of the mass number (top) and the relative fraction of protons, helium and other groups of nuclei to the total CR flux (bottom) obtained in the non-classical diffusion model.}\label{fig:lna}
\end{figure}

The figure~\ref{fig:lna} shows the results of calculations of the mean logarithm of the mass number, as well as the relative contributions of protons, helium and other groups of nuclei to the total CR flux in the energy range of 1~TeV$-$100~PeV. It can be seen evident that the contribution of the light components $p+$He is about 70\%, the CNO group gives $\sim 13$\%. The non-monotonic change in $\ln \langle A \rangle$ is mainly due to a decrease in the contribution of the CNO group in the energy range 0.3$-$3~PeV.

\section{Conclusion}

A new approach to the description of the mechanisms of formation of the features of the spectrum and mass composition of CR in the region of the knee is formulated. The key element of this approach is the CRs non-classical diffusion model developed by the authors. The unique properties of the model~--- the presence of a break in the spectrum, which occurs naturally without using additional assumptions, power-law asymptotics before and after the break, as well as the soft spectrum of particle generation in CR sources~--- are the basis of the new tools for reconstructing the parameters of the sources.

The spectra of the main groups of nuclei were calculated in the energy range of 1~TeV$-$100~PeV within the framework of the non-classical diffusion model. It is shown that the behavior of the all particles spectrum and the mass composition in the knee region is determined by local pevatrons located at the distance of 750$-$900 pc from the Earth. The position of the knee practically coincides with the knees in the spectra of protons and helium nuclei. The contribution of the light components $p+$He is about 70\%, the CNO group gives 13\%. The exponent of the energy spectrum  of the light components is $-$2.61 before the knee. The non-monotonic change in the mean logarithm of mass number is mainly due to a decrease in the contribution of the CNO group in the energy range 0.3$-$3~PeV.

\section*{Funding}

The work is supported by the Russian Science Foundation (grant no. 23-72-00057).

\end{document}